\def\figsize{9.5cm}

\def\rn{}
\def\nn#1 #2{#2. #1}				
\def\nnn#1 #2 #3{#2. #3. #1}			
\def\nnnn#1 #2 #3 #4{#2. #3. #4 #1}		
\def\nnnnn#1 #2 #3 #4 #5{#2. #3. #4 #5. #1}	
\def\dualand{ and\hbox{ }}				
\def\multiand{, and\hbox{ }}				
\def\rf#1;#2;#3;#4;#5 {{\frenchspacing\par\rn#1, #3 {\bf #4}, #5 (#2). \par}}
\def\rrf#1;#2;#3;#4;#5 {{\frenchspacing\rn#1, #3 {\bf #4}, #5 (#2);~}}
\def\rrrf#1;#2;#3;#4;#5 {{\frenchspacing\rn#1, #3 {\bf #4}, #5 (#2).}}
\def\rg#1;#2;#3;#4;#5;#6 {{\frenchspacing\par\rn#1, #3 {\bf #4}, #5 (#2). \par}}
\def\rfbook#1;#2;#3;#4;#5 {{\frenchspacing\par\rn#1, {\it #3} (#5, #4, #2).\par}}
\def\rfprep#1;#2;#3 {{\par\frenchspacing\rn#1, #3 (#2).\par}}
\def\rrfprep#1;#2;#3 {{\frenchspacing\rn#1, #3 (#2);~}}
\def\rrrfprep#1;#2;#3 {{\frenchspacing\rn#1, #3 (#2).}}
\def\rfproc#1;#2;#3;#4;#5;#6 {{\frenchspacing\par\rn#1 #2, in {\it #3}, ed. #4 (#5: #6)\par}}
\def\rfprocp#1;#2;#3;#4;#5;#6;#7 {{\frenchspacing\par\rn#1 #2, in {\it #3}, ed. #4 (#5: #6), p#7\par}}

\def\rg#1;#2;#3;#4;#5;#6 {\par\rn#1 #2, {\it #3}, {\bf #4}, #5 (``#6'') \par}
\def\rf#1;#2;#3;#4;#5 {\par\rn#1, {\it #3}, {\bf #4}, #5 (#2)\par}
\def\rfbook#1;#2;#3;#4;#5 {{\frenchspacing\par\rn#1, {\it #3} (#4: #5, #2)\par}}
\def\rfproc#1;#2;#3;#4;#5;#6 {{\frenchspacing\par\rn#1 #2, in {\it #3}, ed. #4 (#5: #6)\par}}
\def\rfprocp#1;#2;#3;#4;#5;#6;#7 {{\frenchspacing\par\rn#1 #2, in {\it #3}, ed. #4 (#5: #6), p#7\par}}
\def\rfprep#1;#2;#3  {{\par\rn#1, #3, #2\par}}
\def\rfprepp#1;#2;#3 {{\par\rn#1 #2, #3\par}}

\def\etal{{\frenchspacing\it et al.}}
\def\ie{{\frenchspacing\it i.e.}}
\def\eg{{\frenchspacing\it e.g.}}

\def\beq#1{\begin{equation}\label{#1}}
\def\eeq{\end{equation}}
\def\beqa#1{\begin{eqnarray}\label{#1}}
\def\eeqa{\end{eqnarray}}

\def\fig#1{Figure~\ref{#1}}
\def\Fig#1{Figure~\ref{#1}}

\def\spose#1{\hbox to 0pt{#1\hss}}
\def\simlt{\mathrel{\spose{\lower 3pt\hbox{$\mathchar"218$}}
     \raise 2.0pt\hbox{$\mathchar"13C$}}}
\def\simgt{\mathrel{\spose{\lower 3pt\hbox{$\mathchar"218$}}
     \raise 2.0pt\hbox{$\mathchar"13E$}}}
\def\simpropto{\mathrel{\spose{\lower 3pt\hbox{$\mathchar"218$}}
     \raise 2.0pt\hbox{$\propto$}}}

\def\ed{\end{document}}

\def\f{X}
\def\finf{f_\infty}
\def\wi{w_i}
\def\wa{w_a}
\def\wo{w_1}
\def\wz{w_0}
\def\wp{w'_0}

\def\tturn{t_{\rm turn}}
\def\Om{\Omega_m}

\def\Otot{\Omega_{\rm tot}}
\def\zcmb{z_{\rm CMB}}
\def\ztwodf{z_{\rm 2df}}
\def\zmax{z_{\rm max}}

\def\beq#1{\begin{equation}\label{#1}}
\def\eeq{\end{equation}}
\def\beqa#1{\begin{eqnarray}\label{#1}}
\def\eeqa{\end{eqnarray}}

\def\sectio#1{{\bf #1:}}

\documentclass[twocolumn,amsmath,nofootinbib]{revtex4} 
\usepackage{url}
\begin{document}
\input{epsf.sty}




\def\affilmrk#1{$^{#1}$}
\def\affilmk#1#2{$^{#1}$#2;}


\title{New dark energy constraints from supernovae, microwave background \\
and galaxy clustering}

\author{
Yun Wang$^1$ \& Max Tegmark$^{2,3}$
}
\address{
\parshape 1 -3cm 24cm
\affilmk{1}{Department of Physics \& Astronomy, Univ. of Oklahoma, 
440 W.~Brooks St., Norman, OK 73019, USA; wang@nhn.ou.edu}
\affilmk{2}{Department of Physics, University of Pennsylvania,
Philadelphia, PA 19104, USA}
\affilmk{3}{Dept. of Physics, Massachusetts Institute of Technology, 
Cambridge, MA 02139}
}

\date{Submitted to Phys. Rev. Lett. March 11 2004, accepted April 21.}

\begin{abstract}
Using the spectacular new high redshift supernova observations from the HST/GOODS program
and previous supernova, CMB and galaxy clustering data, we make the most accurate measurements to
date of the dark energy density $\rho_X$ as a function of cosmic time, constraining it in a rather 
model-independent way, assuming a flat universe. 
We find that Einstein's vanilla scenario where $\rho_X(z)$ is constant
remains consistent with these new tight constraints,
and that a Big Crunch or Big Rip 
is more than 50 gigayears away for a broader class of models 
allowing such cataclysmic events.
We discuss popular pitfalls and hidden priors: 
parametrizing the equation-of-state $w_X(z)$ 
assumes positive dark energy density and no Big Crunch,
and the popular parametrization
$w_X(z)=w_0 +w_0' z$ has 
nominally strong constraints from CMB
merely because $w_0'>0$ 
implies an unphysical exponential blow-up $\rho_X \propto e^{3 w_0' z}$.
%

\end{abstract}


  
\maketitle


\setcounter{footnote}{0}

The nature of dark energy has emerged as one of the deepest mysteries in physics.
When strong evidence for its existence first appeared from supernova observations
in 1998 \cite{Riess98,Perl99}, the most pressing question was whether it was
real or an observational artifact.
Since then, the supernova evidence has both withstood the test of time and strengthened
\cite{Knop03,Tonry03,Riess04}, and two other lines of evidence
have independently led to the same conclusion: 
measurements of cosmological clustering with the cosmic microwave background (CMB)
and large-scale structure (LSS) (\eg, \cite{Spergel03,sdsspars})
and observation of CMB/LSS correlations due to the late integrated Sachs-Wolfe effect \cite{LISW}.
Now that its current density has been accurately measured 
(WMAP+SDSS gives $\rho_X(0)=(4.8\pm 1.2)\times 10^{-27}$ kg/m$^3$ \cite{sdsspars}, 
corresponding to $(9.3\pm 2.3)\times 10^{-124}$ in Planck units and $\Omega_\Lambda\approx 0.7$),
the next pressing question is clearly whether its density $\rho_X$ 
stays constant over time (like Einstein's cosmological constant)
or varies. The latter is is predicted by most models attempting to explain dark energy
either as a dynamic substance, ``quintessence''
(\eg, \cite{quintessence}),
or via some form of modified gravitational 
theory, perhaps related to extra dimensions or string physics
(\eg, \cite{modifiedgravity})). See \cite{reviews}
for reviews with more complete lists of references.

\begin{figure} 
\vskip-1.3cm
\centerline{\epsfxsize=\figsize\epsffile{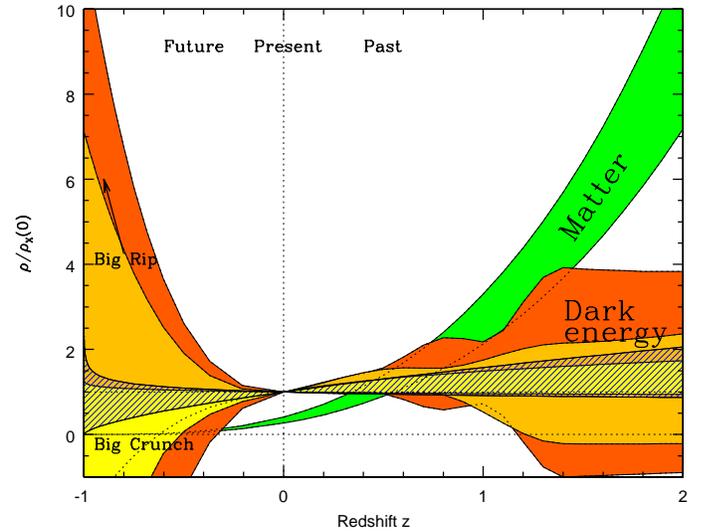}}
\vskip-1.3cm
\caption[1]{\label{SausageFig}\footnotesize%
$1\sigma$ constraints on the density of matter and dark energy 
from SN Ia (Riess sample, flux-averaged with $\Delta z=0.05$), 
CMB and LSS data, all in units of the current 
dark energy density.
From inside out, the four nested dark energy constraints are
for models making increasingly strong assumptions,
corresponding, respectively, to the 4-parameter spline, 
the 3-parameter spline, the 2-parameter $(\finf,\wi)$ case
and the 1-parameter constant $w$ case (hatched).
The Universe starts accelerating when 
the total density slope $d\ln\rho/d\ln(1+z)>-2$, which roughly
corresponds to when dark energy begins to
dominate, \ie, to where the matter and dark energy bands cross.
In the distant future, the Universe recollapses if the dark energy density $\rho_X$
goes negative and ends in a ``Big Rip'' if it keeps growing 
($d\ln\rho_X/d\ln(1+z)<0$).}
\end{figure}

The recent discovery of 16 Type Ia supernovae (SNe Ia) \cite{Riess04} with the Hubble Space 
Telescope during the GOODS ACS Treasury survey bears directly on this question.
By discovering 6 out of the 7 highest-redshift SNe Ia known, all at $z>1.25$, this search 
team \cite{Riess04} was able to 
pinpoint for the first time the transition epoch from matter domination to dark energy
domination when the cosmic expansion began to accelerate.
It is therefore timely to revisit this question of if and how the dark energy density 
varies with time. 
This is the goal of the present paper.
Given our profound lack of understanding of dark energy and the 
profusion of theoretical models in the recent literature, 
we focus on measuring the function $\rho_X(z)$ in as model-independent
a fashion as possible, emphasizing 
what we do and do not know given various 
assumptions about how $\rho_X(z)$ is parametrized, 
about data sets used and about modeling thereof.
We will see that the new data are powerful enough to 
make previous measurements of $\rho_X(z)$ 
(\eg, \cite{Daly03,WangMukherjee03,WangFreese04}) tighter and more robust 
and also to extend them back firmly into the epoch of cosmic deceleration.

\sectio{Analysis Technique}
We wish to measure the dimensionless {\it dark energy function},
$X(z)\equiv\rho_X(z)/\rho_X(0)$,
the dark energy density in units of its present value.
We do this as described in \cite{WangMukherjee03}, fitting 
to SN Ia,
CMB 
and LSS information, obtaining the results shown in Figure 1.


The measured distance-redshift relations of SNe Ia provide the foundation
for probing the dark energy function $\f(z)$. 
In a flat Universe, the dimensionless luminosity distance $d_L(z) H_0/c=(1+z)\Gamma(z)$, where 
$\Gamma(z)=\int_0^z dz'/E(z')$ is the dimensionless comoving distance and
\begin{equation}
E(z) \equiv \left[\Omega_m (1+z)^3 +(1-\Omega_m) \f(z)\right]^{1/2}
\end{equation}
is the cosmic expansion rate relative to its present value.
We use the ``gold'' set of 157 SNe Ia published by Riess {\etal} in \cite{Riess04}
and analyze it using
flux-averaging statistics \cite{Wang00b,WangMukherjee03}
to reduce bias due to weak gravitational lensing by intervening matter. 
We assume spatial flatness as motivated by inflation and discuss the 
importance of this and other assumptions below.
We use CMB and LSS data to help break the degeneracy between the
dark energy function $\f(z)$ and $\Omega_m$. 
For the CMB, we use only the measurement of the CMB shift parameter \cite{Bond97},
$R\equiv \Om^{1/2}\Gamma(\zcmb)=1.716\pm 0.062$ 
from CMB (WMAP, CBI, ACBAR) \cite{Spergel03,CMB},
where $\zcmb=1089$. 
The only large-scale structure information we use is
the linear growth rate $f(\ztwodf)=0.51\pm 0.11$ 
measured by the 2dF galaxy redshift survey (2dFGRS) \cite{2dF,Knop03},
where $\ztwodf=0.15$ is the effective redshift of this survey
and $f\equiv (d\ln D/d\ln a)$ is determined by solving the equation for the 
linear growth rate $D$, 
$D''(\tau) + 2E(z)D'(\tau) - {3\over 2}\Om (1+z)^3D = 0$,
where primes denote $d/d(H_0 t)$.
Note that the CMB and LSS measurements we use ($R$ and $f$) do {\it not} depend on the Hubble parameter $H_0$,
and are quite insensitive to assumptions made about $\f(z)$.
The SN Ia measurements used are also independent of $H_0$, since we marginalize them 
over the intrinsic SN Ia luminosity calibration.

We run a Monte Carlo Markov Chain (MCMC) based on the MCMC engine of \cite{Lewis02}
to obtain a few million samples 
of $\Om$ and $\f(z)$. The dark energy bands in Figure 1 correspond to the
central $68\%$ 
of the $\f$-values at each $z$ and the matter band does the same for 
$\rho_m(z)/\rho_X(0)=(1+z)^3\Om/(1-\Om)$.

\sectio{Results}
\Fig{SausageFig} shows our main results, the constraints on
the dark energy function $\f(z)=\rho_X(z)/\rho_X(0)$
for four different parametrizations, and
illustrates that the assumptions one makes about the curve $\f(z)$
have an important effect on the results.
The most common way of measuring dark energy properties in the literature has been
to parametrize the dark energy function $\f$ by merely one or two free parameters, 
constraining these by fitting to observed data.
Table 1 includes the historically most popular parametrizations, expressed as 
functions of the dimensionless cosmic scale factor $a\equiv (1+z)^{-1}$.
Parametrization A simply assumes that $\f(a)$ is a power law, with the single 
equation-of-state parameter $w$ 
determining its logarithmic slope.
From the identity $\partial\ln\rho_X/\partial\ln a=-3(1+w_x)$, it follows that
parametrization B corresponds to the popular parametrization 
$w_x(z) = \wz + \wp z$ \cite{HutererTurner}, which has been widely used in the literature.
It has the drawback of being rather unphysical for $\wp>0$, with 
the dark energy density $\rho_X(z)$ blowing up as $e^{3\wp z}$ at high redshift.
Parametrization C avoids this \cite{Linder},
and corresponds to $w_x=\wo+\wa(1-a)$,
but blows up exponentially in the future as $a\to\infty$ for $\wa>0$.
In contrast, our 
parametrization D remains well-behaved at all times:
both early on and in the distant future,
the dark energy approaches either a constant equation of state $\wi$ 
or a constant density, depending on the sign of $(1+w_i)$. 

Obviously, the more restrictive the assumptions about $\f$ are, the 
stronger the nominal constraints will be, so it is crucial to be clear on what 
these assumptions are. For instance, Table 1 shows that 
parametrizations A, B and C all tacitly assume that $\f(z)\ge 0$, \ie, that
the dark energy density cannot be negative, hence ruling out by fiat the possibility 
that the Universe can recollapse in a Big Crunch. Note that even 
{\it arbitrary} function $w(z)$ has this hidden assumption built in. 

\begin{table*}
\noindent 
{\footnotesize
Table 1: Parametrizations used for the dark energy function $\f\equiv\rho_X(z)/\rho_X(0)$
in terms of the cosmic scale factor $a=(1+z)^{-1}$.
\begin{center}
\begin{tabular}{|llll|}
\hline
Parametrization			&$n$	&Parameters				&Definition\\
\hline
A) Constant eq.~of state $w$	&1	&$w$					&$\f=a^{-3(1+w)}$\\
B) Affine $w(z)$			&2	&$\wz$, $\wp$			&$\f=a^{-3(1+\wz-\wp)} e^{3\wp(a^{-1}-1)}$\\
C) Affine $w(a)$		&2	&$\wo$, $\wa$				&$\f=a^{-3(1+\wo+\wa)} e^{3\wa(a-1)}$\\
D) Forever regular		&2	&$\wi$, $\finf$				&$\f=\finf + (1-\finf) a^{-3(1+\wi)}$\\
E) 3-parameter spline 		&3	&$\wi$, $\f(z_1)$, $\f(z_2)$		&Cubic spline in $z$ for $z\le z_2$, $\f=\f(z_2)\left({1+z\over 1+z_2}\right)^{3(1+\wi)}$ for $z\ge z_2$\\
F) 4-parameter spline 		&4	&$\wi$, $\f(z_1)$, $\f(z_2)$, $\f(z_3)$	&Cubic spline in $z$ for $z\le z_3$, $\f=\f(z_3)\left({1+z\over 1+z_3}\right)^{3(1+\wi)}$ for $z\ge z_3$\\
\hline
\end{tabular}
\vskip-0.6cm
\end{center}     
} 
\end{table*}

To introduce as little theoretical bias as possible into our measurement,
we use parametrizations E and F from Table 1; these are fairly model-independent 
reconstructions of the dark energy function $\f(z)$, assuming
merely that 
$\f(z)$ is a sufficiently smooth function that it can be modeled with a
cubic spline out to some redshift $\zmax$, and by a constant-$w$ power law 
thereafter. We choose $\zmax$ to avoid sparse SN Ia data,
and parametrize $\f$ by its values at $N$ equispaced
spline points at $\zmax/N$, $2\zmax/N$,...,$\zmax$.
$\f(z)$ is matched smoothly on to $(1+z)^{3(1+w_i)}$ at $z>z_{max}$.
This specifies $\f(z)$ uniquely once we require $\f(z)$ and $\f'(z)$ 
to be everywhere continuous and set $\f(0)=1$, $\f'(0)=\f(z_1)/z_1$.
We have choose $z_{max}=1.4$, as there are only two SNe Ia at 
higher redshifts. Since $\f(z)$ is only very weakly constrained
beyond $z >z_{max}$, we impose a prior of $w_i \geq -2$
to avoid an unbounded parameter space.
Changing the prior to
$w_i \geq -20$ or changing the functional form of $\f(z)$ at $z>z_{max}$
(to an exponential, for example)
has little impact on the reconstructed $\f(z)$.
We also find our results to be rather robost to data details.
Including the ``silver'' sample from \cite{Riess04} does not
change our results qualitatively, and replacing the CMB shift parameter we used
($R=1.716\pm 0.062$) by $R=1.710\pm 0.137$ (from WMAP data alone \cite{Spergel03})
broadens the 68\% 
confidence envelope by less than 20\%.

\Fig{SausageFig} also shows the constraints on the dark energy function
$\f(z)$ corresponding to parametrizations A and D from Table 1,
imposing the priors $w_i \geq -2$ and $\finf\geq 0$ for D.
For comparison
with the results of \cite{Riess04},
we also studied 
parametrization B,
with a weak prior $\wp\geq -20$ to avoid an unbounded parameter space.
Note that MCMC tacitly assumes uniform prior on the
parameters, so if the parameter space is unbounded, the MCMC will 
drift off in the unbounded direction and never converge.
Reparametrizing changes this implicit prior
by the Jacobian of the transformation.
Although we have imposed minimal priors to avoid unbounded parameter space
where $X(z)$ can be arbitrarily close to zero, but we have {\it not}
imposed priors motivated by any theoretical model.
For example, scalar-field models typically have
$\f'(a)\le 0$, since fields usually roll down potentials, not up.
In addition, many models prohibit the dark energy density from being negative. 
However, we do not wish to assume such priors, since 
``dark energy'' could be a manifestation of something completely different, like modified 
gravity \cite{modifiedgravity}. 


As has been emphasized \cite{WangGarnavich01,gravity,Maor02},
SN Ia data are sensitive only to the smooth, overall shape of $\f(z)$.
This is because the error bars on sharp features on a scale $\Delta z$ 
are proportional to $(\Delta z)^{-3/2}$ due to the derivative involved in 
going from comoving distance $r(z)$ to dark energy function $\f$ \cite{gravity} 
--- reconstructing $w_X(z)$ is still
harder, the requirement that one effectively take the second derivative of noisy data
\cite{WangFreese04} giving the error scaling as $(\Delta z)^{-5/2}$ \cite{gravity}.
\Fig{SausageFig} shows that as we allow more small-scale freedom by parametrizing $\f(z)$ by 
1, 2, 3 and 4 parameters, the allowed bands become thicker. However, 
the broader bands generally encompass the narrower ones, showing 
no hint in the data that the true $\f(z)$ has funny features outside of the
1- and 2-parameter model families. Indeed, all bands are seen to be consistent
with the simplest model of all: the zero-parameter ``vanilla'' model 
$X(z)=1$ corresponding to Einstein's cosmological constant.

In other words, faced with the fact that an analysis using parametrization A 
implies $w\approx -1$ (we obtain $w=-0.91^{+0.13}_{-0.15}$ combining SN Ia, CMB and LSS),
readers hoping for something more interesting than vanilla may correctly argue that
these constraints are dominated by accurate measurements at lower redshift and may fail
to reveal hints of an upturn in $\f(z)$ at $z\simgt 1$ because parametrization A 
incorrectly assumes that $(\log a,\log\f)$ is a straight line.
Our more general parametrizations close this loophole
by allowing $\f(z)$ much greater freedom,
and the fact that none of them provide any hint yet of non-vanilla dark energy behavior
therefore substantially strengthens the case for a simple cosmological constant, $\f(z)=1$.

What is the ultimate fate of the Universe? 
If for any of our models $\rho_X$ eventually goes negative so that total density drops to zero
at some time $\tturn$, then the expansion reverses and a Big Crunch occurs at 
$t=2\tturn$ --- this applies only if $\f$ is uniquely determined by the cosmic scale factor 
(equivalently $z$) as in Table 1, and not for many scalar field models \cite{Kallosh}.
The cosmic time $t = \int da/\dot a = \int H^{-1}d\ln a$, 
and if this asymptotes to a finite value as $a\to\infty$, then a cataclysmic 
Big Rip \cite{Caldwell03} occurs at this time. This is equivalent to  $w(z)<-1$ at $z=-1$,
so parametrizations A, B and C rip if $w<-1$, $\wz-\wp<-1$ and $\wa>0$, respectively.
 
Predictions for the future need to be taken with a large grain of salt, since 
they are obviously highly model-dependent. 
For instance, parametrizations A, B and C cannot crunch, 
whereas E and F cannot rip.
Simply combining all MCMC models from all our parametrizations, 
we find that 95\% 
of them last at least another 49 gigayears,
25\% 
ending in a Big Crunch,
8\% 
ending in a Big Rip
and 67\%
quietly expanding forever. 
%
%
  
\sectio{Caveats and potential pitfalls}
When interpreting dark energy constraints such as those that we have presented, 
two crucial caveats must be borne in mind: 
potential SN Ia systematic errors and potential false assumption about other physics.
We refer the reader to \cite{Knop03,Riess04} for thorough discussions of the former
and focus on the latter.

The SN Ia, CMB and LSS measurements we have used involve 
only $\f(z)$, $\Om$ and $\Otot$.
Because of degeneracies between these three quantities,
the inferences about $\f(z)$ therefore depend strongly on the assumptions
about the two cosmological parameters $\Om$ and $\Otot$.
Yet it is all too common to constrain dark energy properties using 
prior information about $\Om$ and $\Otot$ that
in turn hinges on assumptions about the dark energy, usually the vanilla assumption $\f(z)=1$,
a pitfall emphasized by, \eg, \cite{Maor02}.

We have assumed flat space, $\Otot=1$, as have virtually all recent publications
measuring dark energy properties (usually using parametrizations A, B or C).
It is well-known that this assumption is crucial: introducing $\Otot$ as a free parameter
to be marginalized over has such a dramatic effect on luminosity distances that essentially
no interesting constraints can be placed on $\f(z)$ at the present time, not even 
assuming the highly restrictive parametrization A.
We will present a detailed investigation of dark energy independent constraints 
on $\Otot$ from CMB and LSS elsewhere.

\begin{figure} 
\vskip-0.3cm
\centerline{\epsfxsize=\figsize\epsffile{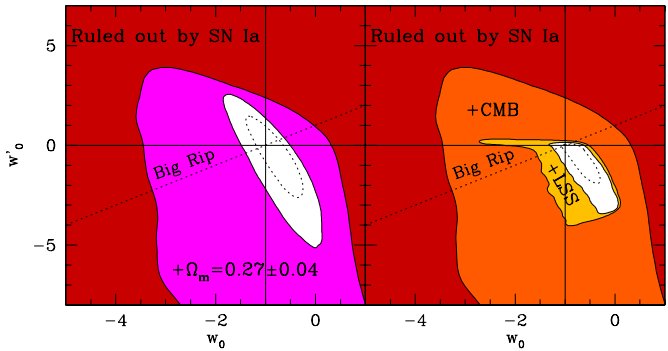}}
\vskip-4.5cm
\caption[1]{\label{wwpFig}\footnotesize%
How constraints on $w_0$ and $w_0'$
depend on assumptions and data used. Darker shaded regions are ruled
out at 95\% confidence by SNe Ia alone; lighter shaded regions
are ruled out when adding other information
as indicated. 68\% 
contours are dotted. Models above the dotted line end in a Big Rip.
The 157 SNe Ia (Riess sample) have been flux-averaged with $\Delta z=0.05$.
}
\vskip-4mm
\end{figure}


We now turn to the issue of dark-energy independent constraints on $\Om$.
As emphasized by \cite{Maor02}, assumptions about $\Om$ make a 
crucial difference as well. As an example, \fig{wwpFig} shows the constaints
on $(\wz,\wp)$ for parametrization B. The left panel illustrates
that the constraints from SN Ia alone are much weaker than those 
obtained by imposing a strong prior $\Om=0.27\pm 0.04$ as was done 
in Figure 10 of \cite{Riess04}.
Although this prior coincides with the measurement of $\Om$ from 
WMAP and 2dFGRS \cite{Spergel03}, it should {\it not} be used here
since it assumes $\f(z)=1$.
%
%
The right panel of \Fig{wwpFig} shows the 
effect of including CMB information self-consistently (via the $R$-parameter)
in our constraints.
We see that $\wz$-values as low as $-3$ remain allowed, as expected 
given the above-mentioned weak $\Om$-constraints, and that additional 
information (in this case from LSS) is needed to tighten things up.
This panel also illustrates the hazard of poor dark energy parametrizations:
the seemingly impressive upper limit on $\wp$ tells us nothing whatsoever
about dark energy properties via SN Ia, but merely reflects that the
unphysical exponential blowup  $\f\propto e^{3\wp z}$ would violate
the CMB constraint.

\sectio{Conclusions}
In conclusion, we have reported the most accurate measurements to date of
the dark energy density $\rho_X$ as a function of time,
assuming a flat universe.
We have found that in spite of their constraining power, the 
spectacular new high-$z$ supernova measurements 
of \cite{Riess04} provide no hints of departures from the
vanilla model corresponding to Einstein's cosmological constant.
This is good news in the sense of simplifying the rest of cosmology,
but dims the prospects that
nature will give us quantitative clues about the true nature of dark energy by
revealing non-vanilla behavior. The apparent constancy of $\rho_X(z)$
also makes attempts to explain away dark energy by blaming systematic errors appear
increasingly contrived, further strengthening the evidence that 
dark energy is real and hence a worthy subject of study.
Future experiments \cite{FutureSN} can dramatically shrink the error bars in 
Figure 1, and therefore hold great promise for illuminating the nature of dark energy.

{\bf Public software:} A Fortran code that uses flux-averaging statistics 
to compute the likelihood of an arbitrary 
dark energy model (given the SN Ia data from \cite{Riess04})  
can be found at $http://www.nhn.ou.edu/\sim wang/SNcode/$.

{\bf Acknowledgements:}
We thank Adam Riess for sending us the data of 
a low-$z$ SN Ia missing from the preprint of \cite{Riess04}, and
Dragan Huterer, Jan Kratochvil, Andrei Linde, Eric Linder, Pia Mukherjee, 
H\aa vard Sandvik and Paul Steinhardt for helpful discussions.
This work was supported by
NSF CAREER grants AST-0094335 (YW) and AST-0134999 (MT), 
NASA grant NAG5-11099 and fellowships from the David and Lucile
Packard Foundation and the Cottrell Foundation (MT).




\vskip-0.7cm
\begin{thebibliography}{99}


\bibitem{Riess98}
\rf\nnn Riess A G {\etal};1998;Astron. J.;116;1009

\bibitem{Perl99} 
\rf\nn Perlmutter S {\etal};1999;ApJ;517;565

\bibitem{Knop03}
\rf\nnn Knop R A {\etal};2003;ApJ;598;102

\bibitem{Tonry03}
\rf\nnn Tonry J L {\etal};2003;ApJ;594;1 

\bibitem{Riess04}
\rfprep\nnn Riess A G {\etal};2004;astro-ph/0402512


\bibitem{Spergel03}
\rf\nnn Spergel D N {\etal} (WMAP);2003;ApJS;148;175 

\bibitem{sdsspars}
M. Tegmark {\etal}, {astro-ph/0310723, PRD, in press}


\bibitem{LISW}  
\rrf\nn Boughn S\dualand\nn Crittenden R;2004;Nature;427;45
\rrfprep\nn Nolta M {\etal};2003;astro-ph/0305097
\rrfprep\nn Fosalba P\dualand\nn Gaztanaga E;2003;astro-ph/0305468
\rrf\nn Fosalba P, \nn Gaztanaga E\multiand\nn Castander F;2003;ApJ;597;L89
\rrfprep\nn Scranton R {\etal};2003;astro-ph/0307335
\rrrfprep\nn Afshordi N, \nn Loh Y\multiand\nnn Strauss M A;2003;astro-ph/0308260

\bibitem{quintessence}
\rrf\nn Freese K {\etal};1987;Nucl.Phys.;B287;797
\rrf\nnnn Peebles P J E\dualand\nn Ratra B;1988;ApJ;325;L17
\rrf\nn Wetterich C;1988;Nucl.Phys.;B302;668 
\rrf\nnn Frieman J A, \nnn Hill C T, \nn Stebbin A\multiand\nn Waga I;1995;PRL;75;2077 
\rrf\nn Caldwell R, \nn Dave R\multiand\nnn Steinhardt P J;1998;PRL;80;1582

\bibitem{modifiedgravity}  
\rrf\nn Parker L\dualand\nn Raval A;1999;PRD;60;063512
\rrf\nn Deffayet C;2001;Phys.Lett.B;502;199 
\rrf\nn Mersini L, \nn Bastero-Gil M\multiand\nn Kanti P;2001;PRD;64;043508
\rrf\nn Freese K\dualand\nn Lewis M;2002;Phys.Lett.B;540;1 

\bibitem{reviews}
\rrf\nn Padmanabhan T;2003;Phys.Rep.;380;235 
\rrrf\nnnn Peebles P J E\dualand\nn Ratra B;2003;Rev.Mod.Phys.;75;559

\bibitem{Daly03}
\rrf\nnn Daly R A\dualand\nnn Djorgovski S G;2003;ApJ;597;9
\rrfprep\nn Alam U, \nn Sahni V, \nnn Saini T D\multiand\nn Starobinsky A A;2003;astro-ph/0311364
\rrfprep\nnn Choudhury T R\dualand\nn Padmanabhan T;2003;astro-ph/0311622 

\bibitem{WangMukherjee03}
Y. Wang Y and P. Mukherjee; {astro-ph/0312192}

\bibitem{WangFreese04}
\rfprep\nn Wang Y\dualand\nn Freese K;2004;astro-ph/0402208

\bibitem{Caldwell03}
\rf\nnn Caldwell R R, \nn Kamionkowski M\multiand\nnn Weinberg N N;2003;Phys.Rev.Lett.;91;071301
      
\bibitem{Wang00b}
\rf\nn Wang Y;2000;ApJ;536;531


\bibitem{CMB}
\rrf\nnn Pearson T J {\etal} (CBI);2003;ApJ;591;556 
\rrf\nnn Kuo C L {\etal} (ACBAR);2004;ApJ;600;32  

\bibitem{Bond97}
\rf\nnn Bond J R, \nn Efstathiou G\multiand\nn Tegmark M;1997;MNRAS;291;L33

\bibitem{2dF}
\rrf\nn Hawkins E {\etal};2003;MNRAS;346;78
\rrf\nn Verde L {\etal};2002;MNRAS;335;432 

\bibitem{Lewis02}
\rf\nn Lewis A\dualand\nn Bridle S;2002;PRD;66;103511


\bibitem{Linder}
\rf\nn Linder E;2003;Phys. Rev. Lett.;90;091301 




\bibitem{WangGarnavich01}
\rf\nn Wang Y\dualand\nnn Garnavich P M;2001;ApJ;552;445 


\bibitem{gravity}
\rfprep\nn Tegmark M;2001;astro-ph/0101354

\bibitem{Maor02}
\rrf\nn Maor I, \nn Brustein R\multiand\nnn Steinhardt P J;2002;PRL;86;6 
\rrf\nn Maor I, \nn Brustein R, \nn McMahon J\multiand\nnn Steinhardt P J;2002;PRD;65;123003 


\bibitem{FutureSN}
\rf\nn Wang Y;2000;ApJ;531;676


\bibitem{HutererTurner}
\rrf\nn Huterer D\dualand\nnn Turner M S; 2001;PRD;64;123527

\bibitem{Kallosh}
\rfprep\nn Kallosh R {\etal};2003;astro-ph/0307185
      


\end{thebibliography}
\end{document}